\documentclass[pra,aps,superscriptaddress,groupedaddress,showpacs,twocolumn,amsmath,amssymb]{revtex4}

\usepackage{graphicx}
\usepackage{amssymb}
\usepackage{color}
\usepackage{psfrag}
\usepackage{ifsym}
\usepackage{hyperref}
\usepackage[hyperpageref]{backref}

\definecolor{pink}{rgb}{1,0.078,0.57}

\definecolor{green}{rgb}{0,0.7,0.9}

\newcommand{\ket}[2] {| #1 \rangle_{#2}}
\newcommand{\bra}[2] {\langle #1 |_{#2}}

\newcommand{\dg}{^{\dagger}}

\begin{document}

\title{Optimization of quantum state tomography in the presence of experimental constraints}

\author{Mohammadreza Mohammadi}
\affiliation{Department of Physics, University of Toronto, 60 Saint George St., Toronto, Ontario M5S 1A7, Canada}
\author{Agata M. Bra\'nczyk }
\email{abranczyk@perimeterinstitute.ca}
\affiliation{Department of Physics, University of Toronto, 60 Saint George St., Toronto, Ontario M5S 1A7, Canada}
\affiliation{Department of Chemistry, University of Toronto, 80 Saint George St., Toronto, Ontario M5S 3H6, Canada}
\affiliation{Perimeter Institute for Theoretical Physics, Waterloo, Ontario N2L 2Y5, Canada}

\date{\today}

\begin{abstract}
In the absence of experimental constraints, optimal measurement schemes for quantum state tomography are well understood. We consider the scenario where the experimenter doesn't have arbitrary freedom to construct their measurement set, and may therefore not be able to implement a known optimal scheme. We introduce a simple procedure for minimizing the uncertainty in the reconstructed quantum state for an arbitrary tomographic scheme. We do this by defining a figure of merit based on the equally-weighted variance of the measurement statistics.  This figure of merit is straightforwardly based on the singular value decomposition of the measurement matrix, making it well-suited for optimization. 
\end{abstract}


\maketitle

The characterization of quantum states is a key step in quantum technologies such as quantum computing \cite{Kok2007}, quantum information \cite{Nielsen2010} and quantum cryptography \cite{Bennett1984}. This can be achieved by performing an informationally complete set of measurements on multiple identically-prepared copies of the state, using the results to reconstruct a representation of the state. Such a procedure is known as \emph{quantum state tomography} (QST) \cite{Vogel1989,Leonhardt1995,James2001}. 

In principle, as long as the measurement set is informationally complete, the density matrix representing the state can always be reconstructed (Sec. \ref{sec:standardtomo}). However, in practice, certain measurement schemes are more prone to error than others. 

The optimality of various tomographic schemes has been explored in works such as those by Scott \cite{Scott2006}, Roy and Scott \cite{Roy2007}, and de Burgh \emph{et al.} \cite{deBurgh2008}. Roughly speaking, for two-state systems, measurement schemes  whose probability operators are distributed symmetrically on the Bloch sphere---such as those based on mutually unbiased bases and platonic solids---minimize the error in the reconstructed density matrix. 

In this paper, we consider a scenario where the experimenter lacks the resources to implement an arbitrary measurement set, and may therefore not be able to implement a known optimal scheme. We show how one can determine the measurement settings which  minimize the uncertainty in the reconstructed quantum state. 

To do so, we introduce a figure of merit based on the equally-weighted variance (EWV) of the measurement statistics, that can be used to quantify the uncertainty in the reconstructed density matrix (Sec \ref{sec:noisy}). This figure of merit is based on the singular-value decomposition of the measurement matrix and is trivial to compute on a modern computer, making it ideal for the task of optimization. We demonstrate this figure of merit's utility (Sec \ref{sec:noisyFTT}), by optimizing wave-plate parameters in two non-standard optical tomographic scenarios. 

%%%%%%%%%%%%%%%%%%%
\section{Tomographic protocol}\label{sec:standardtomo}
%%%%%%%%%%%%%%%%%%%

In a typical tomography experiment, multiple copies of the same quantum state interact with a measurement apparatus. The apparatus determines the number of states that registered the outcome $j$, known as `counts'. 

This can be modelled with an informationally complete set of probability operators $\hat{E}_{j}$. For our purposes, it is more natural to work directly with measurement statistics rather than probabilities. We therefore consider \emph{unnormalized} probability operators that form an \emph{unnormalized} POVM  such that $\sum_{j}\hat{E}_{j}=\mathcal{N}\mathbb{I}$, where $\mathcal{N}$ is the total number of counts for the POVM (for a nice introduction to the POVM formalism, refer to Ch 2.2.6 of \cite{Nielsen2010}). In the absence of an informationally complete POVM, multiple POVMs can be combined to form an informationally complete set. 

The number of counts for a given outcome $j$ is given by the expectation value of the corresponding probability operator
\begin{eqnarray}\label{eq:2}
{n}_{j}=\langle \hat{E}_{j}\rangle= \mathrm{Tr}[\hat \rho \hat{E}_{j}]\,,
\end{eqnarray} 
where $\hat\rho$ is the density matrix of a $d$-level quantum system. The total number of counts for a single POVM is given by $\mathcal{N}=\sum_{j}{n}_{j}$.

Given a set of measurement outcomes ${n}_{j}$, one is interested in inferring the density matrix of the measured quantum state. This may be done by first decomposing the unknown density matrix $\hat\rho$ in terms of an orthogonal basis:
\begin{eqnarray} \label{eq:1}
\hat \rho=\frac{1}{\sqrt {2d} }  \sum_{{i}=0}^{d^{2}-1} {S}_{i} \hat \sigma_{i}    ,
\end{eqnarray}
where $\hat\sigma_{0}=\sqrt{2/d}~\mathbb{I} $ is related to the identity matrix $\mathbb{I}$ and $\hat \sigma_{i}$ are the generalized Pauli matrices---a set of $d$-dimensional traceless, Hermitian matrices that are orthogonal under the Hilbert-Schmidt inner product, given by $\mathrm{Tr}[\hat\sigma_{i},\hat\sigma_{j}]=2\delta_{{i}{j}}$. The parameters ${S}_{i}$ completely characterize the state $\hat\rho$ and the condition ${S}_{0} = 1$ ensures that the density matrix is normalized.

Inserting Eq.  (\ref{eq:1}) into Eq. (\ref{eq:2}), we find a linear relationship between the parameters ${S}_{i}$ and the measurement outcomes ${n}_{{j}}$:
\begin{eqnarray}\label{eq:matN}
{n}_{{j}}=\frac{1}{\sqrt {2d} } \sum_{{i}=0}^{d^{2}-1} {S}_{i}  \mathrm{Tr}[\hat \sigma_{i}  \hat{E}_{{j}}]\,.
\end{eqnarray} 
This can be written in compact matrix form as
\begin{align}\label{eq:matP}
\mathbf{n}=\mathbf{M}\mathbf{a}\,,
\end{align}
where the vectors
\begin{align} 
\mathbf{n}={}&(n_{1}, \dots ,n_{N})^{\mathrm{T}}\,;\\
\mathbf{a}={}&({S}_{0}, {S}_{1},\dots,{S}_{d^2-1} )^T\,,
\end{align}
are related by the measurement matrix
\begin{align}\label{eq:column}
\mathbf{M}_{i, j}= \frac{1}{\sqrt{2d}}\mathrm{Tr}[ \hat \sigma_{j-1} \hat{E}_{i}] \,.
\end{align}

If the tomography scheme consists of  $N\geq {d}^2$ operators and is informationally complete (i.e. $\mathbf{M}$ has ${d}^2$ singular values), the density matrix of an unknown state $\hat\rho$ can be reconstructed by calculating the pseudo-inverse $\mathbf{M}^+$ of the measurement matrix $\mathbf{M}$  to give the parameters $\mathbf{a}$ in terms of the measurement outcomes $\mathbf{n}$:
\begin{align}\label{eq:matPinv}
\mathbf{a}=\mathbf{M}^+\mathbf{n}\,.
\end{align}
If $N=d^2$, $\mathbf{M}^+=\mathbf{M}^{-1}$ is simply the inverse of $\mathbf{M}$ and if $N> {d}^2$, the tomography scheme may lead to an overcomplete set of Eqs. for ${n}_{j}$. 

A specific example of how the above formalism can be used to implement the familiar six-outcome tomographic protocol for qubits can be found in Appendix \ref{sec:example}.

We note that the inversion technique described above amounts to performing least-squares estimation which does not always produce physical density matrices. Despite this, we argue in Appendix \ref{sec:pos} that linear inversion provides a reasonable starting point for developing a figure of merit that characterizes the uncertainty in the reconstructed density matrix.

When it comes to actual state reconstruction in practice, one often resorts to using the popular maximum likelihood estimation method \cite{Hradil1997}. Alternatively, one can look to a growing number of exciting new techniques such as the forced purity routine \cite{Kaznady2009}, Baysean mean estimation \cite{Blume-Kohout2010}, hedged maximum likelihood estimation \cite{Blume-Kohout2010a}, compressed sensing \cite{Gross2010a}, von Neumann entropy maximization \cite{Teo2011}, minimax estimation \cite{Ng2012} and likelihood-free quantum inference \cite{Ferrie2013a}, as well as techniques that focus on reconstructing the state with reliable error bars \cite{Christandl2011} and confidence regions \cite{Blume-Kohout2012}. 
 
\section{Deriving a figure of merit}\label{sec:noisy}

Our goal is to identify the set of operators $\{\hat{E}_{i}\}$ that minimize the uncertainty in the reconstructed density matrix. To do so, we must first define a figure of merit that quantifies this uncertainty. 

In our analysis, we consider statistical uncertainties in measurement outcomes rather than systematic errors, e.g. due to misalignment of optical elements. We therefore neglect error in $\mathbf{M}$. We note that a recent study comparing systematic and statistical errors in QST was performed by Langford \cite{Langford2013}. 

Under these assumptions, We can relate uncertainties  in the detected number of counts to uncertainties in the density matrix by
\begin{align}\label{eq:matPinvD}
\Delta\mathbf{a}=\mathbf{M}^+\Delta\mathbf{n}\,,
\end{align}
where
\begin{align}
\Delta\mathbf{n}={}&(\Delta {n}_{1},\dots,\Delta {n}_{N})^{\mathrm{T}}\,;\\
\Delta\mathbf{a}={}&(\Delta{S}_{0}, \Delta{S}_{1}, \dots ,\Delta{S}_{d^{2}-1}  )^T\,.
\end{align}

To assess a tomography scheme's robustness to noise, we calculate the equally-weighted variance $\mathrm{EWV}$, defined in terms of the equal sum of the variances in the parameters $\mathbf{a}$  \cite{Sabatke2000}:
\begin{align}
\mathrm{EWV}=\sum_{i=0}^{d^{2}-1}{\Delta {S}_{i}^2}=\Delta\mathbf{a}\cdot\Delta\mathbf{a}\,.
\end{align}
From Eq. (\ref{eq:matPinvD}), it follows that 
\begin{align}
\mathrm{EWV}={}&({\mathbf{M}}^+\Delta{\mathbf{n}})\cdot({\mathbf{M}}^+\Delta{\mathbf{n}})\\
={}&\sum_{i=1}^{d^{2}}\Big(\sum_{j=1}^{N} {\mathbf{M}}^{+}_{{i}{j}} \Delta{n}_{j}\Big)^2\,.
\end{align}

Statistical noise can take the form of signal-independent background noise such as additive Gaussian noise, or signal-dependent noise such as Poisson noise. The latter is endemic to experiments involving particles that are spontaneously generated, such as single photons or neutrons. 

Here, we consider Poisson noise, which leads to a figure of merit that is input-state-dependent. An average over all input states gives an expression that corresponds to the result for signal-independent noise. 

Fluctuations due to Poisson noise are statistically independent from one measurement to the other \cite{Goudail2009}, simplifying the expression to
\begin{align}
\mathrm{EWV}={}&\sum_{i=1}^{d^{2}}\sum_{j=1}^{N} \left({\mathbf{M}}^{+}_{{i}{j}}\right)^2 (\Delta{n}_{j})^2\,.
\end{align}
For Poisson processes, the  variance is given by $(\Delta{n}_{j})^2={n}_{j}$, thus
\begin{align}
\mathrm{EWV}={}& \sum_{i=1}^{d^{2}}\sum_{j=1}^{N} \left({\mathbf{M}}^{+}_{{i}{j}}\right)^2{n}_{j}\,.
\end{align}
From Eq. (\ref{eq:matP}), we see that ${n}_{j}=\sum_{k}\mathbf{M}_{jk}\mathbf{a}_{k}$, giving
\begin{align}\label{eq:something}
\begin{split}
\mathrm{EWV}={}& \sum_{{i},{k}=1}^{d^{2}}\sum_{j=1}^{N} \left(\mathbf{M}^{+}_{{i}{j}}\right)^2\mathbf{M}_{jk}\mathbf{a}_{k}\,.
\end{split}
\end{align}

In Appendix \ref{sec:simpI}, we make use of the singular value decomposition (SVD) $\mathbf{M}^{+}=(\mathbf{V}^{+})^T\mathbf{S}^{+}\mathbf{U}^{+}$ (i.e.  $\mathbf{M}=\mathbf{U}\mathbf{S}\mathbf{V}^{\mathrm{T}}$ ) to arrive at the final expression for the equally-weighted variance:
\begin{align}\label{eq:im}
\begin{split}
\mathrm{EWV}={}& \sum_{{i}=1}^{d^2}\frac{1}{\mu_{i}^2}\sum_{{k}=1}^{d^{2}}\mathbf{r}_{{i},{k}}\mathbf{a}_{k}\,,
\end{split}
\end{align}
where $\mu_{i}$ are the singular values of $\mathbf{M}$ and
\begin{align}\label{eq:r}
\mathbf{r}_{{i},{k}}={}&\sum_{j=1}^{d^2}(\mathbf{U}_{{i}{j}}^{-1})^2\mathbf{M}_{{j}{k}}=\sum_{j=1}^{d^2}\mathbf{U}_{{j},{i}}^2\mathbf{M}_{{j}{k}}\,.
\end{align}

Eq. (\ref{eq:im}) is an expression for the equally-weighted variance of a tomography scheme undergoing Poisson noise statistics. As expected for signal-dependent noise processes, it is dependent on the target state given by the vector $\mathbf{a}$. In the next section, we calculate the average equally-weighted variance. 

We emphasize that our figure of merit is applicable for tomographic schemes described by the above formalism. Namely, those where the unknown state is parametrized by a Bloch vector; all Bloch vector elements are of equal interest; and that the probability operators form a POVM. Take for counter-example the scheme for measuring the diagonal elements of the density matrix of a single-mode state of an electromagnetic field introduced by Mogilevtsev \cite{Mogilevtsev1998}. For this scheme, using our notation, we have $\hat\sigma_{i}=\ket{i}{}\bra{i}{}$ and $\hat{E}_{j}=\sum_{k}(1-\eta_{j})^{k}\ket{k}{}\bra{k}{}$, in the Fock basis. The measurement matrix is then given by the Vandermonde matrix. Naive calculation of the EWV from this measurement matrix may not lead to reasonable assessment of the uncertainty in this scheme, since $\hat\sigma_i$ do not form a complete basis and $\hat{E}_{j}$ do not satisfy the condition $\sum_{j}\hat{E}_{j}=\mathcal{N}\mathbb{I}$.

\subsection{Average EWV}
In many cases, one is interested in how well a tomography scheme performs overall, rather than for one particular state.  In this section, we address this question by calculating the average equally-weighted variance for all possible input states:
\begin{align}\label{eq:avg1}
\begin{split}
\langle\mathrm{EWV}\rangle={}& \sum_{{i}=1}^{d^2}\frac{1}{\mu_{i}^2}\sum_{{k}=1}^{d^{2}}\mathbf{r}_{{i},{k}}\langle\mathbf{a}_{k}\rangle \,.
\end{split}
\end{align}

The average over the first element of $\mathbf{a}$ is trivial, i.e. $\langle \mathbf{a}_{1}\rangle= \langle S_{0} \rangle=1 $. To calculate the average for other values of $k$, i.e. $\left < {S}_{k} \right > ~\forall~k\neq 0$, we follow the approach of \.Zyczkowski \emph{et al.}  \cite{Zyczkowski2011} and assume an average over an ensemble of random states for which the probability measure may be factorized. The distribution of eigenvalues and eigenvectors are therefore independent and, hence, can be averaged independently. While the average over eigenvectors is relatively straightforward, performing the average over eigenvalues can be quite tricky as there is no single probability measure for mixed states. Fortunately, as we show below, for any set of eigenvalues, the average over all possible eigenvectors gives $\left < {S}_{k} \right >=0 ~\forall~k\neq 0$, saving us from the difficulty of averaging over eigenvalues. 

To calculate $\left < {S}_{k} \right >$ we invert Eq. (\ref{eq:1}) such that 
\begin{align} \label{eq:component}
 \mathbf{a}_{k+1} = S_{k} = \sqrt {\frac{d}{2}}\mathrm{Tr}\Big[ \sum_{i=1}^{d}\lambda_{i}\ket{\psi_{i}}{}\bra{\psi_{i}}{} \rho \hat \sigma_{k}\Big]\,,
\end{align}
where we used $\mathrm{Tr}[\hat\sigma_{i}\hat\sigma_{j}]=2\delta_{{i}{j}}$ and made use of the eigenvalue decomposition $\hat \rho = \sum_{i=1}^{d}\lambda_{i}\ket{\psi_{i}}{}\bra{\psi_{i}}{}$ where $\lambda_{i} \geq{} 0$ and $\sum_{i=1}^{d}\lambda_{i}={}1$. We can write the average of Eq. (\ref{eq:component}) over all possible eigenbases $\langle\cdot\rangle_{\mathrm{b}}$ for a given set of eigenvalues $\{\lambda_i\}$ as
\begin{align} \label{eq:avg}
\left < {S}_{k} \right >_{\mathrm{b}} ={}&  \sqrt {\frac{d}{2}} \Big <\mathrm{Tr}\Big[ \sum_{i=1}^{d}\lambda_{i}\ket{\psi_{i}}{}\bra{\psi_{i}}{}  \hat \sigma_{k}\Big] \Big >_{\mathrm{b}}  \,.
\end{align}

As pointed out in  \cite{Zyczkowski2011}, it is natural to assume that the eigenvectors are distributed according to the unique, unitarily invariant, Haar measure on unitary $d \times d$ matrices.
Different assignments of eigenvectors to eigenvalues are all equally likely since any rearrangement of the eigenvectors could be done by a unitary transformation, and the Haar measure is invariant under unitary transformations. For a given set of eigenvalues $\{\lambda_i\}$, we can therefore decompose the sum over eigenbases into a sum over permutations of $\{1,2, \dots, d \}$ as follows
\begin{align} \label{eq:avg}
\left < {S}_{k} \right >_{\mathrm{b}} ={}&   \frac{\sqrt{d}}{\sqrt{2}d!} \Big < \mathrm{Tr}\Big[ \sum_{i=1}^{d} \sum_{\pi \in P_{d}}\lambda_{i}\ket{\psi_{\pi(i)}}{}\bra{\psi_{\pi(i)}}{}  \hat \sigma_{k}\Big] \Big >_{\mathrm{b}} \,,
\end{align}
where  $P_{d}$ is the set of all possible maps that permute $\{1,2, \dots, d \}$. We can rewrite the sum over permutations of eigenbases in Eq. (\ref{eq:avg}) as the sum over permutations of eigenvalues as follows
\begin{align} 
\left < {S}_{k} \right >_{\mathrm{b}} 
={}&\frac{\sqrt{d}}{\sqrt{2}d!}\Big < \mathrm{Tr}\Big[ \sum_{i=1}^{d} \left ( \sum_{\pi \in P_{d}}\hspace{-0.2cm}\lambda_{\pi^{-1}(i)} \right )\ket{\psi_{i}}{}\bra{\psi_{i}}{}  \hat \sigma_{k}\Big] \Big >_{\mathrm{b}} \,,
\end{align}
where $\pi^{-1}$ is the inverse of the map $\pi$. We note that $\sum_{\pi \in P_{d}}\hspace{-0.2cm}\lambda_{\pi^{-1}(i)}= \sum_{\pi \in P_{d}}\lambda_{\pi(i)}$ and that this is independent of the index $i$. We can therefore write

\begin{align} \label{eq:avg}
\left < {S}_{k} \right >_{\mathrm{b}} ={}& \frac{\sqrt{d}}{\sqrt{2}d!} \left ( \sum_{\pi \in P_{d}}\lambda_{\pi(i)} \right ) \Big < \mathrm{Tr}\Big[ \sum_{i=1}^{d} \ket{\psi_{i}}{}\bra{\psi_{i}}{}  \hat \sigma_{k}\Big] \Big >_{\mathrm{b}}  \,.
\end{align}
Using the fact that $\sum_{i=1}^{d}\ket{\psi_{i}}{}\bra{\psi_{i}}{}= \mathbb {I}$, and that the generalized Pauli matrices are traceless, we find that $\left < {S}_{k} \right >_{\mathrm{b}}=0 ~\forall~k\neq 0$. We therefore conclude that $\left < {S}_{k} \right >=0 ~\forall~k\neq 0$,  and  Eq. (\ref{eq:avg1}) simplifies to
\begin{align}\label{eq:avg2}
\begin{split}
\langle\mathrm{EWV}\rangle={}& \sum_{{i}=1}^{d^2}\frac{1}{\mu_{i}^2}\mathbf{r}_{{i},{1}} \,.
\end{split}
\end{align}
Inserting the expression for $\mathbf{M}_{j, i}$ in Eq. (\ref{eq:column}) into the expression for $\mathbf{r}_{{i},{k}}$ in (\ref{eq:r}), we evaluate the expression for $\mathbf{r}_{{i},{1}}$:
\begin{subequations}
\begin{align}
\mathbf{r}_{{i},{1}}={}&\sum_{j=1}^{d^2}\mathbf{U}_{{j},{i}}^2 \mathbf{M}_{j, 1}\\
={}& \sum_{j=1}^{d^2} 
 \frac{1}{\sqrt{2d}}\mathbf{U}_{{j},{i}}^2 \mathrm{Tr}[ \hat \sigma_{0} \hat{E}_{j}] \,.
\end{align}
\end{subequations}

The operators $\hat{E}_{j}$ can be thought of as scaled projectors such that  $\mathrm{Tr}[ \hat{E}_{j}]=\alpha_{j}$. We use this, as well as the definition  $\hat\sigma_{0}=\sqrt{2/d}~\mathbb{I} $,  to give
\begin{align}
\mathbf{r}_{{i},{1}}={}&\frac{1}{d}\sum_{j=1}^{d^2}\alpha_{j} \mathbf{U}_{{j},{i}}^2 \,,
\end{align}
which combined with Eq. (\ref{eq:avg2}) gives
\begin{align}\label{eq:im3}
\left < \mathrm{EWV} \right >={}& \frac{1}{d}  \sum_{{{i}{j}}=1}^{d^2} \frac{\alpha_{j} \mathbf{U}_{{j},{i}}^2} {\mu_{i}^2} \,.
\end{align}
Eq. (\ref{eq:im3}) is an expression for the average EWV of a tomography scheme undergoing Poisson noise statistics. This expression is trivial to compute using a modern computer and does not require computational averaging over input states---in contrast to, say, the fidelity \cite{deBurgh2008}---making it ideal for optimization over tuneable parameters.

If all probability operators are equally weighted, i.e. $\alpha_{j}=\alpha$, Eq. (\ref{eq:im3}) simplifies to
 \begin{align}\label{eq:equalalpha}
\left < \mathrm{EWV} \right >={}& \frac{\alpha}{d}   \sum_{{i}=1}^{d^2} \frac{1} {\mu_{i}^2} \, ,
\end{align}
since each column of the unitary matrix $\mathbf{U}$ has unit length.

We note that the $\langle \mathrm{EWV}\rangle$ corresponds to the EWV for signal-independent noise, such as additive Gaussian noise. 

\subsection {Lower bound for $\left < \mathrm{EWV} \right >$}
When evaluating the robustness of a specific tomography scheme to noise, it is useful to make comparisons with the best possible scheme, i.e. one that minimizes $\left < \mathrm{EWV} \right >$.

In this section we derive a lower bound for the average EWV. The approach we use is to find a relationship between the diagonal elements and singular values of the Hermitian matrix $\mathbf{M}^{\mathrm{T}}\mathbf{M}$. This puts constraints on the parameters which determine $\langle\mathrm{EWV}\rangle$. We then use Lagrange multipliers, which provide a strategy for optimizing a function subject to equality constraints, to minimize $\langle\mathrm{EWV}\rangle$.

First, consider the singular value decomposition of the matrix $\mathbf{M}^{\mathrm{T}}\mathbf{M}$, where $\mathbf{M}$ is  defined in Eq. (\ref{eq:matP}). 
\begin{subequations}
\begin{align}\label{eq:113}
\mathbf{M}^{\mathrm{T}} \mathbf{M} ={}&\left(\bf{V}\mathbf{S^{\mathrm{T}}}\bf{U}^{\mathrm{T}}\right)\left(\bf{U}\mathbf{S}\bf{V}^{\mathrm{T}}\right)\\\label{eq:B2}
={}&\bf{V}\mathbf{S^{\mathrm{T}}} \mathbf{S}\bf{V}^{\mathrm{T}}\\\label{eq:B3}
={}&\bf{V}\bf{W}\bf{V}^{\mathrm{T}}\,,
\end{align}
\end{subequations}
where $\bf{V}$ is a $d^{2}\times d^{2}$ orthogonal matrix. Comparing Eqs. (\ref{eq:B2}) and (\ref{eq:B3}), we find that the eigenvalues $\{ \lambda_{i}\} _{i=1}^{d^{2}} $ of the matrix $\mathbf{M}^{\mathrm{T}} \mathbf{M}$ are related to the singular values of the matrix $\mathbf{M}$ by $\lambda_{i}=\mu_{i}^2$.

To calculate a lower bound for the uncertainty, we begin by considering the sum of the diagonal elements  $\{d_{i}\}_{i=1}^{d^{2}}$ of the matrix  $\mathbf{M}^{\mathrm{T}} \mathbf{M} $: 
\begin{align}
\sum_{j=1}^{d^{2}} d_{j}={}&\sum_{i=1}^{N}\sum_{j=1}^{d^{2}} \mathbf{M}_{j,i}^T\mathbf{M}_{{i}{j}}\\\label{eq:dd}
={}&\frac{1}{2d} \sum_{i=1}^{N}\sum_{j=0}^{d^{2}-1} \mathrm{Tr}[ \hat \sigma_{j} \hat{E}_{i}] ^{2}. 
\end{align}
The set of operators
\begin{align}\label{eq:basis}
 \mathbf{B} = \Big\{ \frac{\hat \sigma_{j}}{\sqrt{2}}    \Big\}~~\mathrm{for}~~ j=0,\dots, d^{2}-1
\end{align} 
forms an orthonormal basis, in the sense of Hilbert-Schmidt inner product, for the vector space of all  $d \times d$ Hermitian matrices over real numbers. Recall that $\hat{E}_{i}$ are scaled projectors with a trace $\alpha_{i}$. We can therefore write the ``Hilbert-Schmidt length'' of the vector $\hat{E}_{j}$ in this vector space as
\begin{align} \label{eq:diagonal}
\mathrm{Tr}[\hat{E}_{i} \hat{E}_{i}]=  \frac{1}{2} \sum_{{j}={0} } ^{d^{2}-1}\mathrm{Tr}[\hat \sigma_{j} \hat{E}_{i}]^{2} =\alpha_{i}^2\,,
\end{align}
which simplifies Eq. (\ref{eq:dd}) to
\begin{align}\label{eq:dd2}
\sum_{j=1}^{d^{2}} d_{j}={}&\frac{\sum_{i}^{N}\alpha_{i}^{2}}{d}\,.
\end{align}
We now make use of the Schur-Horn theorem \cite{Schur1923,Horn1954}, which states that there exists an ${m}\times {m}$ Hermitian matrix with  diagonal values $\{d_{i}\} _{i=1}^{m} $ and eigenvalues $\{ \lambda_{i}\} _{i=1}^{m} $ that are both ordered non-increasingly, if and only if, 
\begin{align}
\sum_{i=1}^{l} d_{i} \leq \sum_{i=1}^{l} \lambda_{i}, \> \forall l \in \{1, 2, ..., {m} \}
\end{align}
and
\begin{align}
\sum_{i=1}^{m} d_{i} = \sum_{i=1}^{m} \lambda_{i}\,. 
\end{align}

Bearing in mind that the matrix $\mathbf{M}^{\mathrm{T}} \mathbf{M} $ is Hermitian, we find that 
\begin{subequations}\label{eq:b6}
\begin{align}
\sum_{i=1}^{d^{2}}\mu_{i}^{2}={}&\sum_{j=1}^{d^{2}} d_{j}={}\frac{\sum_{i}^{N}\alpha_{i}^{2}}{d}\,;\\
\mathrm{and}~~~~~~~\mu_{1}^{2}={}&\lambda_{1} \geq \frac{\sum_{i}^{N}\alpha_{i}^{2}}{d^2}\,.
\end{align}
\end{subequations}

Given the above constraints and the fixed sum $\sum_{i}^{N}\alpha_{i} $, which corresponds to having a fixed total number of counts for a given POVM, we used Lagrange multipliers to show that for a given $\mu_{1}$, the expression $ \sum_{k=1}^{d^2} 1/\mu_{k}^{2}$ is minimal when $\mu_{2}=\dots=\mu_{d^2}$ and $\alpha:=\alpha_{1}=\dots=\alpha_{N}$. This reduces the problem to a single-variable optimization problem, which yields
\begin{subequations}\label{eq:singlenoise}
\begin{align}\label{eq:minimumsingular}
\mu_{1}^{2}={}&\frac{N\alpha^2}{d^{2}}\,;\\  \label{eq:minimumsingular2}
\mathrm{and}~~~~~~~\mu_{2}^{2}=\dots={}&\mu_{d^{2}}^{2} = \frac{N\alpha^{2}}{d^{2}(d+1)} \,.
\end{align}
\end{subequations}
Under these conditions, we can find a lower bound for the average EWV by inserting the expressions for $\mu_{i}$ in Eqs. (\ref{eq:singlenoise}) into  Eq. (\ref{eq:equalalpha}). The final expression for the lower bound for the average EWV is
\begin{subequations}\label{eq:im4}
 \begin{align}
\left < \mathrm{EWV} \right >_{\mathrm{LB}}={}&\frac{d^{2}(d^{2}+d-1)}{N\alpha}  \,.
\end{align}
\end{subequations}
Given a completely unknown state, one is interested in a tomographic scheme that performs best ``on average''.
Symmetric informationally-complete positive operator valued measures (SIC-POVMs) are known to be optimal for quantum state tomography under signal-independent noise \cite{Scott2006}. We find that they are also optimal given Poisson noise processes, by showing that the $\left < \mathrm{EWV} \right >$ for SIC-POVMs is equal to $\left < \mathrm{EWV} \right >_{\mathrm{LB}}$ (see Appendix \ref{sec:sicpovm}).

\section{Examples}\label{sec:noisyFTT}

The $\left < \mathrm{EWV} \right >$ can be used to determine the optimal tomography scheme under a set of practical constraints. We demonstrate this with two examples, using two-state systems, i.e. qubits, encoded in the polarization degree of freedom of single photons. In such systems, the measurement is typically made in the horizontal/vertical basis by counting photons at the output ports of a polarizing beam splitter (PBS). Measurements in different bases are implemented by placing wave plates at different orientations in front of the beam splitter. 

\begin{figure}[t]
\includegraphics[width=0.6\columnwidth]{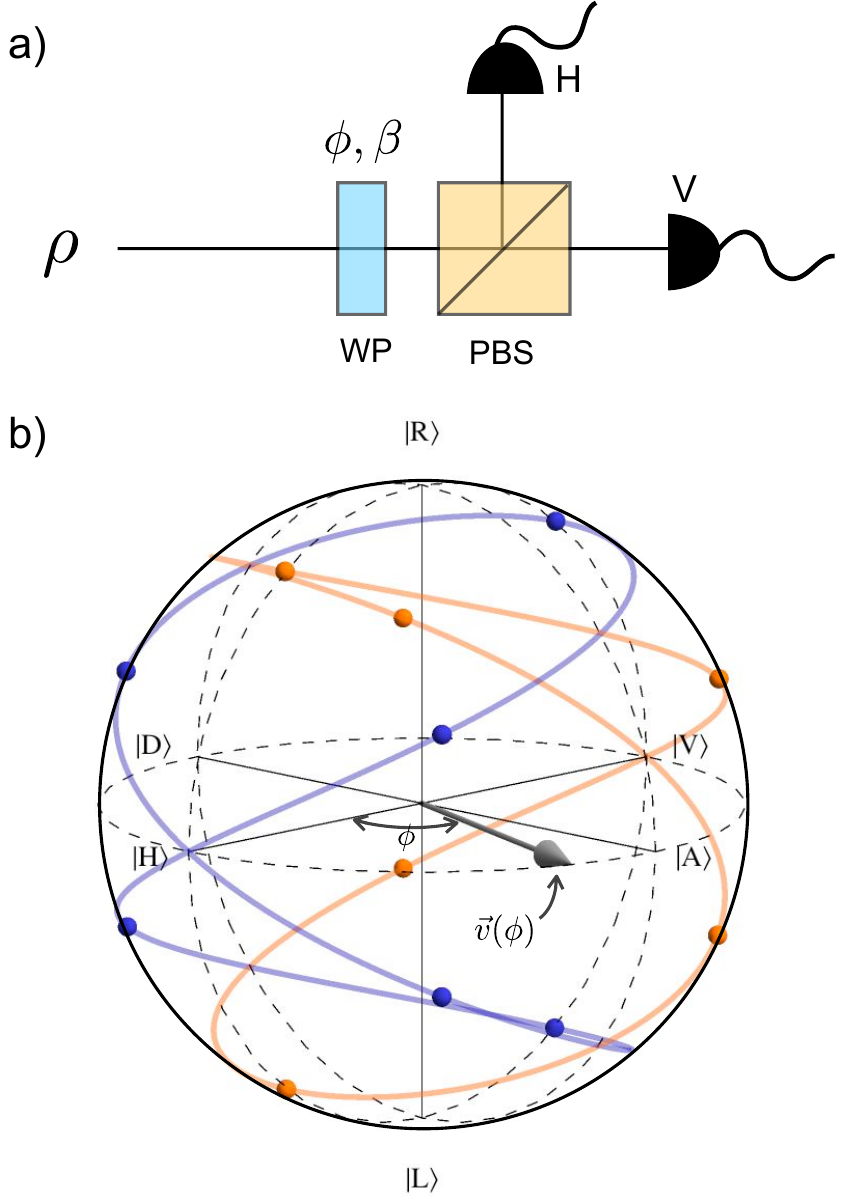}
\caption{ (Color online)  a) Schematic diagram of the Fourier Transform Tomography set-up with one rotating wave plate \cite{Mohammadi2013}. The angle $\phi$ parameterizes the orientation of the wave plate about the optical axis. b) Path traced out by $\hat{E}_{H}(\beta,\phi)$ (blue) and $\hat{E}_{V}(\beta,\phi)$ (orange) defined in Equation (\ref{eq:MeasurementOperator}), for: $\beta= 7\pi/10$  and $N=12$. }
\label{fig:Fig1}
\end{figure}

In Scenario A, we are restricted to one wave plate per qubit mode, but have the freedom to choose the retardance of each wave plate (Fig. \ref{fig:Fig1} a). We use $\left< \mathrm{EWV} \right >$ to optimize over the retardance of the wave plate. 

In Scenario B, we consider two wave plates with which we can perform QST on a single qubit mode, however, the wave plates are optimized for a different optical frequency to the one used in the experiment (Fig. \ref{fig:Fig3} a). We minimize $\left< \mathrm{EWV} \right >$ to optimize over the orientation of the wave plates about the optical axis.

%%%% FIGURE %%%%
%%%%%%%%%%%%%
\subsection{Optimization of wave plate retardance}
A single wave plate in front of a PBS can generate a POVM consisting of two operators associated with the horizontal and vertical output modes of each PBS, given by $\{\hat{E}_{H}(\beta,\phi),\hat{E}_{V}(\beta,\phi)\}$ where
\begin{align}\label{eq:MeasurementOperator}
\hat{E}_{a}(\beta,\phi)={}&\mathcal{N}\hat{U}\dg(\beta,\phi)\ket{a}{}\bra{a}{}\hat{U}(\beta,\phi)\,,
\end{align}
for ${a}=H,V$.  In fact, these operators correspond to the special case of a projective-value measure (PVM), where $\hat{E}_i\hat{E}_{j}=\hat{E}_i\delta_{i,j}$.  The unitary operator associated with the wave plate,
\begin{eqnarray}\label{eq:Unitary}
 \hat U(\beta,\phi)= \cos \left (\frac {\beta}{2}\right )\hat \sigma_{0} -i\sin \left (\frac{\beta}{2}\right )\vec {v}(\phi) \cdot \vec{\sigma}  \,,
\end{eqnarray}
rotates the operators $\ket{a}{}\bra{a}{}$ on the Bloch sphere by an angle $\beta$, about the vector 
\begin{align}
\vec {v}(\phi)=\cos(\phi) \vec k+\sin(\phi) \vec i\,,
\end{align}
where $\vec k$ and $\vec i$ are unit vectors in Euclidian space (defined by the axes in Fig. \ref{fig:Fig1}) and $\vec {v} \cdot \vec{\sigma} = {v}_{1}\hat \sigma_{1}+{v}_{2}\hat \sigma_{2}+{v}_{3}\hat \sigma_{3}$. As the wave plate is rotated about the optical axis, the projector $\hat{E}_{H}(t)$ (after appropriate normalization) traces out a figure-eight path on the Bloch sphere, as shown in Fig. \ref{fig:Fig1}. The retardance of the wave plate $\beta$ determines the size of the figure eight. 

Although $\{\hat{E}_{H}(\beta,\phi),\hat{E}_{V}(\beta,\phi)\}$ is not informationally complete, an informationally-complete set of operators can be constructed with different orientations of the wave plate about the optical axis, given by the angle $\phi$, as long as the retardance is not equal to an integer multiple of $\pi$. Here, the total number of projectors $N$ is twice the number of PVMs used to construct the complete tomographic protocol. This scenario was considered in the \emph{Fourier Transform Tomography} (FTT) scheme introduced in \cite{Mohammadi2013}.

The task is to identify the optimal retardance $\beta$ that minimizes the average EWV. We consider a tomographic protocol which consists of a set of PVMs whose elements are distributed along the figure eight with equally-spaced values of $\phi$. From these operators, we calculate the coefficients $\alpha_{i}=\mathrm{Tr}[\hat{E}_{i}]=\mathcal{N}$.  We also construct a measurement matrix $\mathbf{M}$ according to Eq. (\ref{eq:column}) and calculate its singular values $\mu_{i}$. For $N=12$ (i.e. six different values of the retardance which corresponds to six different PVMs), the measurement matrix is given by:
\begin{align}
\mathbf{M}^{(12)}=\frac{\mathcal{N}}{2}\left(
\begin{array}{cccc}
 1 & \frac{- \sqrt{3}(\cos \beta -1)}{4}  & \frac{-\sin \beta }{2} & \frac{(\cos \beta +3)}{4}    \\
 1 & 0 & -\sin \beta  & \cos \beta  \\
 1 & \frac{ \sqrt{3}(\cos \beta -1)}{4}  & \frac{-\sin \beta }{2} & \frac{(\cos \beta +3)}{4}    \\
 1 & \frac{- \sqrt{3}(\cos \beta -1)}{4}  & \frac{\sin \beta }{2} & \frac{(\cos \beta +3)}{4}    \\
 1 & 0 & \sin \beta  & \cos \beta  \\
 1 & \frac{ \sqrt{3}(\cos \beta -1)}{4}  & \frac{\sin \beta }{2} & \frac{(\cos \beta +3)}{4}    \\
 1 & \frac{ \sqrt{3}(\cos \beta -1)}{4}  & \frac{\sin \beta }{2} &  \frac{-(\cos \beta +3)}{4}
    \\
 1 & 0 & \sin \beta  & -\cos \beta  \\
 1 & \frac{- \sqrt{3}(\cos \beta -1)}{4}  & \frac{\sin \beta }{2} &  \frac{-(\cos \beta +3)}{4}
    \\
 1 & \frac{ \sqrt{3}(\cos \beta -1)}{4}  & \frac{-\sin \beta }{2} &  \frac{-(\cos \beta +3)}{4}
    \\
 1 & 0 & -\sin \beta  & -\cos \beta  \\
 1 & \frac{- \sqrt{3}(\cos \beta -1)}{4}  & \frac{-\sin \beta }{2} &  \frac{-(\cos \beta +3)}{4}
\end{array}
\right)
\end{align}

%%%% FIGURE %%%%
\begin{figure}[t]
\includegraphics[width=0.8\columnwidth]{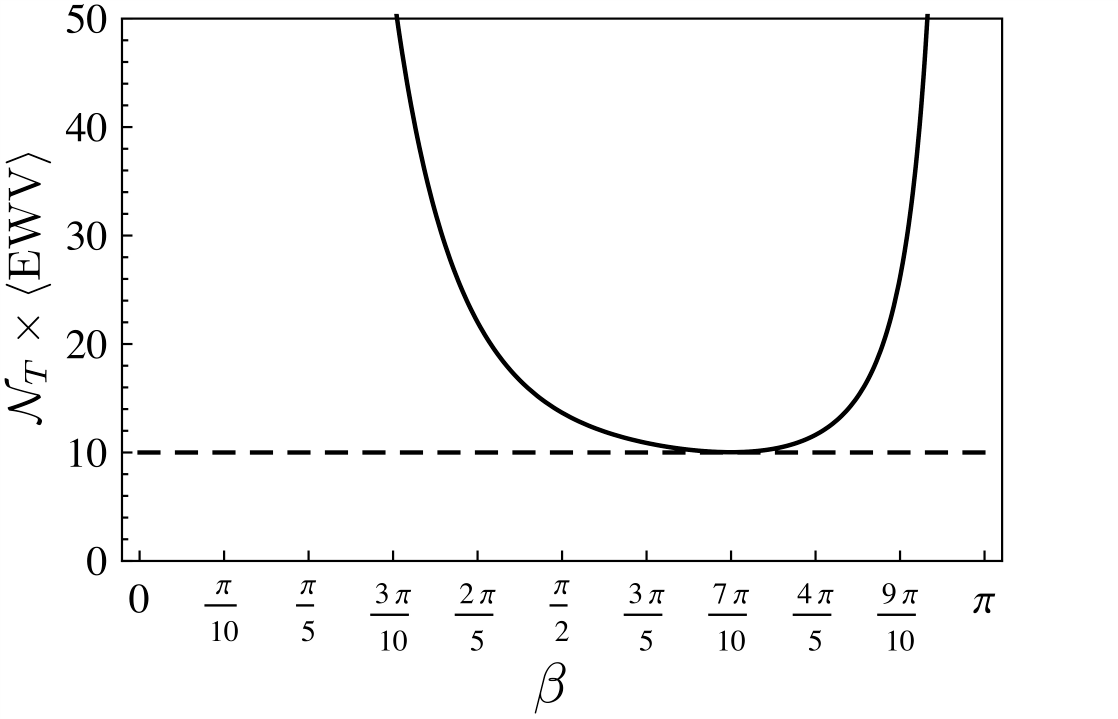}
\caption{ The average equally weighted variance $\left < \mathrm{EWV} \right >$ for tomography with one wave plate, scaled by the total number of counts $\mathcal{N}_{T}$, as a function of the retardance $\beta$ for $N=12$. The dotted line shows the lower bound $\left < \mathrm{EWV} \right >_{\mathrm{LB}}=10/\mathcal{N}_{T}$, attainable by optimal tomographic schemes such as those that measure the Pauli matrices or the SIC-POVM \cite{Rehacek2004,Ling2006,Kalev2012a}. Note that the EWV is dimensionless.}
\label{fig:Fig2}
\end{figure}
%%%%%%%%%%%%%

Using Mathematica \cite{Wolfram-Research2010}, the singular values were found to be:
\begin{subequations}
\begin{align}
\mu_{1}={}&\frac{\mathcal{N}\sqrt{N}}{\sqrt{2}} \,;\\
\mu_{2}={}& \frac{ \mathcal{N}\sqrt{N} }{4\sqrt{2}}   \sqrt{9+4 \cos (\beta )+3 \cos (2 \beta )}\,;\\
\mu_{3}={}&\frac{ \mathcal{N}\sqrt{N}}{2}   |\sin (\beta )| \,;\\
\mu_{4}={}&\frac{ \mathcal{N}\sqrt{N}}{2}    \sin^2\left(\frac{\beta }{2}\right)\,,
\end{align}
\end{subequations}
for $N\geq 5$. Knowing $\mu_{i}$ and $\alpha_{i}$, we can calculate the average EWV according to Eq. (\ref{eq:im3}), which is given by
\begin{align}
\begin{split}
\left < \mathrm{EWV} \right >={}&\frac{2}{ \mathcal{N}_{T}}\Big(\frac{1}{2}+ \csc ^4\left(\frac{\beta
   }{2}\right)+ \csc ^2(\beta )\\
   &+\frac{8}{9+4\cos(\beta)+3\cos(2\beta)}\Big)\,,
\end{split}
\end{align}
where $\mathcal{N}_{T}= \mathcal{N}N/2$ is the total number of counts. Fig. \ref{fig:Fig2} shows $\left < \mathrm{EWV} \right >$ as a function of the retardance $\beta$ for 6 equally-spaced time bins for a single qubit (i.e. $N=12$). This function reaches a minimum of $\left < \mathrm{EWV} \right >\approx10.03/\mathcal{N}_{T}$ when $\beta\approx 7\pi/10$. For comparison, a protocol that consists of three  PVMs (see Appendix \ref{sec:example})---where the measurement operators of each PVM are the eigenstates of the Pauli operators---gives $\left < \mathrm{EWV} \right >=10/\mathcal{N}_{T}$, where $\mathcal{N}_{T}=3\mathcal{N}$. The four-outcome SIC-POVM also gives $\left < \mathrm{EWV} \right >=10/\mathcal{N}_{T}$, where here $\mathcal{N}_{T}=\mathcal{N}$.

%%%% FIGURE %%%%
\begin{figure}[t]
\includegraphics[width=0.6\columnwidth]{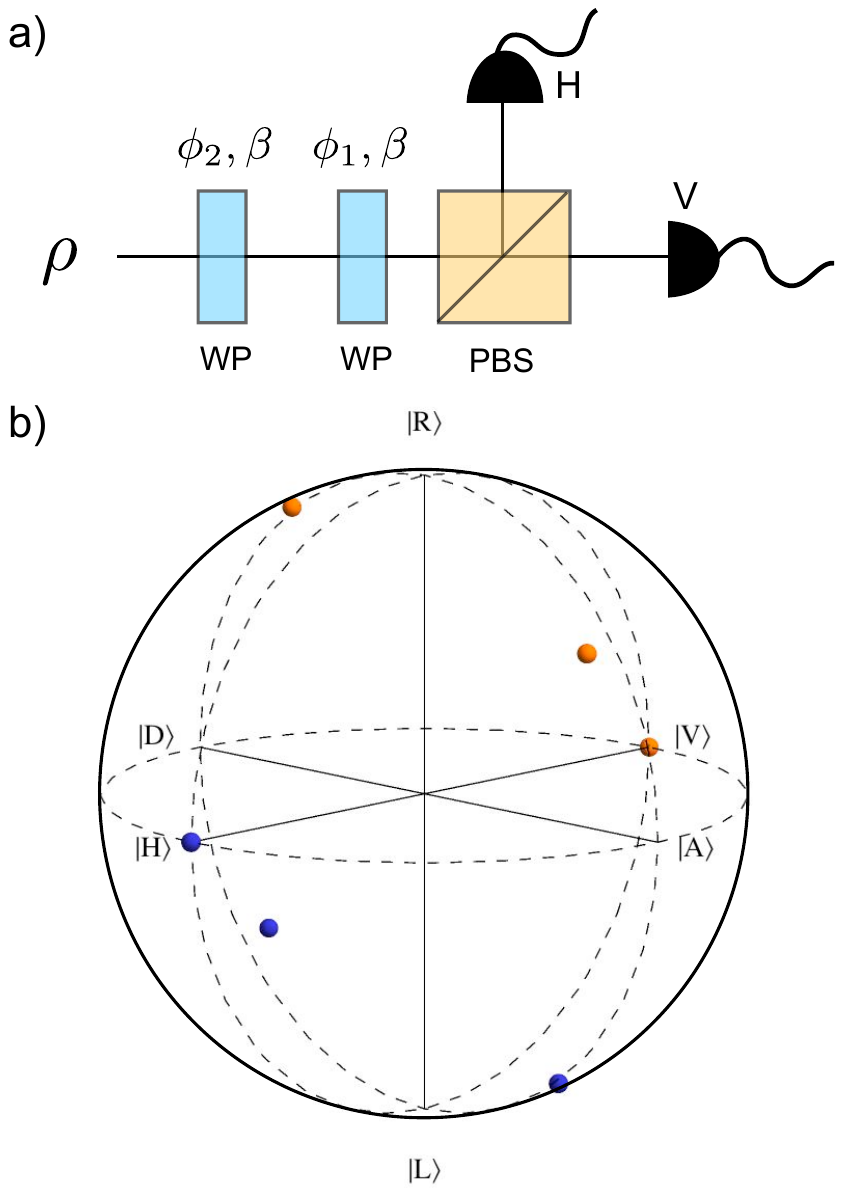}
\caption{ (Color online)  a) Schematic diagram of a tomography set-up with two wave plates of retardance $\beta$. The angles $\phi_1$ and $\phi_2$ parameterize the orientation of the wave plates about the optical axis. b) The six projectors  that minimize the $\left < \mathrm{EWV} \right >$; $\{\hat{E}_{a}(\tfrac{3\pi}{8},0,0),\hat{E}_{a}(\tfrac{3\pi}{8},\tfrac{7\pi}{10},\tfrac{7\pi}{10}),\hat{E}_{a}(\tfrac{3\pi}{8},\tfrac{7\pi}{10},\tfrac{\pi}{5})\}$ where $a=H,V$ (see Eq. (\ref{eq:MeasurementOperatorx})).}
\label{fig:Fig3}
\end{figure}
%%%%%%%%%%%%%

%%%% FIGURE %%%%
\begin{figure}[b]
\includegraphics[width=0.8\columnwidth]{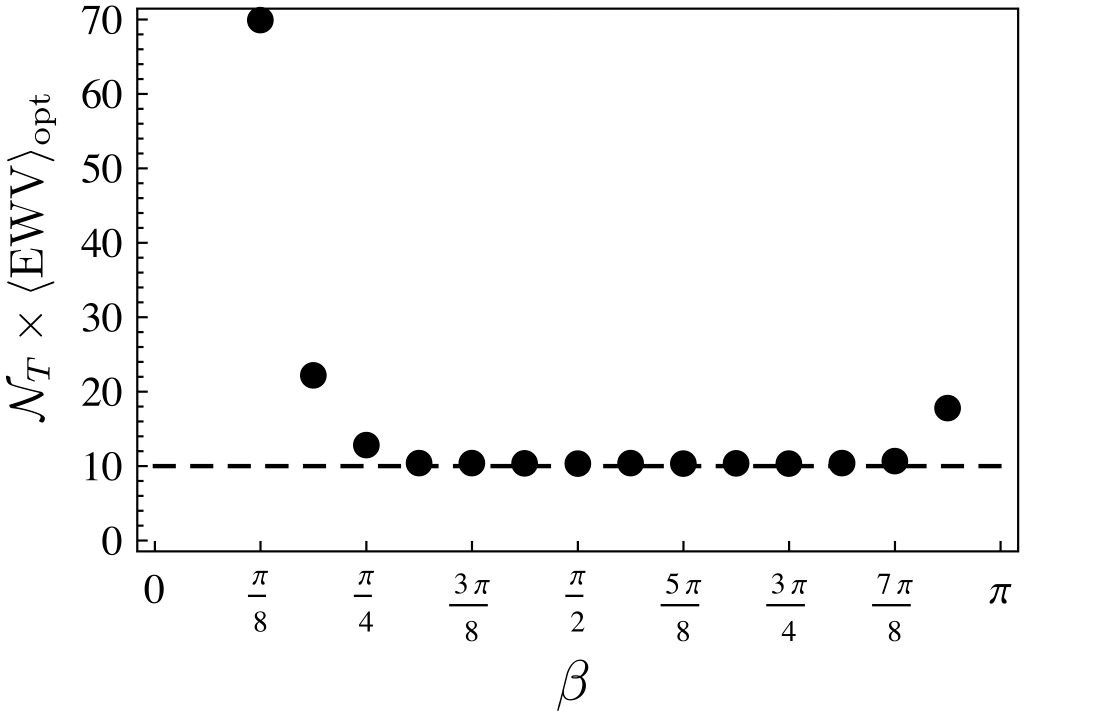}
\caption{ a) The optimal average equally weighted variance $\left < \mathrm{EWV} \right >_{\mathrm{opt}}$ for tomography with two identical wave plates, scaled by the total number of counts $\mathcal{N}_{T}$, for different values of the retardance $\beta$ . The dotted line shows the lower bound $\left < \mathrm{EWV} \right >_{\mathrm{LB}}=10/\mathcal{N}_{T}$. Note that the EWV is dimensionless. Note that unlike in Fig. \ref{fig:Fig2}, the minimization was performed numerically for discrete values of $\beta$. }
\label{fig:Fig4}
\end{figure}
%%%%%%%%%%%%%

\subsection{Optimization of wave plate orientation}

The second scenario we consider consists of optimization over wave plate orientation. Given two wave plates---a half-wave plate ($\beta=\pi/2$) and a quarter-wave plate ($\beta=\pi/4$)---one is able to access the entire surface of the Bloch sphere. A wave plate's retardance, however, is dependent on the wavelength of the input light; using the wrong wave plate results in a non-standard retardance. Here we consider the situation where the experimenter only has access to two identical wave plates, designed for a different optical wavelength. In such a scenario, access to the entire Bloch sphere may not be possible. 

For two wave plates of retardance $\beta$, the measurement operators associated with the  horizontal and vertical output modes of each PBS are given by $\{\hat{E}_{H}(\beta,\phi_1,\phi_2),\hat{E}_{V}(\beta,\phi_1,\phi_2)\}$ where
\begin{align}\label{eq:MeasurementOperatorx}
\begin{split}
\hat{E}_{a}(\beta&,\phi_1,\phi_2)\\
={}&\mathcal{N}\hat{U}\dg(\beta,\phi_{2})\hat{U}\dg(\beta,\phi_{1})\ket{a}{}\bra{a}{}\hat{U}(\beta,\phi_{1})\hat{U}(\beta,\phi_{2})\,,
\end{split}
\end{align}
for ${a}=H,V$, where $\hat{U}(\beta,\phi)$ is defined in Eq. (\ref{eq:Unitary}). 

To construct an informationally-complete set of operators, we generate three PVMs from (\ref{eq:MeasurementOperatorx}) using different realizations of $\phi_1$ and $\phi_2$. We fix $\phi_{1}=\phi_{2}=0$ for the first PVM and optimize over $\phi_{1}$ and $\phi_{2}$ for the second and third PVMs.  We follow the same procedure as above---constructing a measurement matrix $\mathbf{M}$ to find its singular values $\mu_{i}$ as well as calculating $\alpha_{i}$--- to compute $\left < \mathrm{EWV} \right >$.

For $\beta=3\pi/8$, optimization over $\phi_{1}$ and $\phi_{2}$ yields $\left < \mathrm{EWV} \right >\approx10.01/\mathcal{N}_{T}$, which occurs at $\phi_1\approx7\pi/10$ and $\phi_2\approx7\pi/10$ for the second PVM, and  $\phi_1\approx7\pi/10$ and $\phi_2\approx\pi/5$ for the third PVM. The resulting six projectors are shown in Fig \ref{fig:Fig3} b). Fig \ref{fig:Fig4} shows the optimal $\left < \mathrm{EWV} \right >$ for different $\beta$; showing that there exists a large retardance window that can lead to effectively optimal state reconstruction.

\section{Concluding remarks}\label{sec:conc}

We have shown how one can use a simple figure of merit to optimize a tomographic protocol when practical constraints prohibit the implementation of known optimal protocols. In doing so, we considered statistical errors in the form of Poisson noise and assumed no systematic errors. One could in principle follow a similar procedure for noise models appropriate to other experimental scenarios, although whether such a simple figure of merit arrises is to be determined. Systematic errors could also be included by considering uncertainties in the measurement matrix, e.g. $\Delta \mathbf{a}=\mathbf{M}+\Delta\mathbf{n}+\Delta\mathbf{M}+\mathbf{n}$.

This procedure could be straightforwardly extended to quantum process tomography \cite{Chuang1997a,Altepeter2003,OBrien2004} where the measurement statistics relate linearly to the unknown density matrix parameters. Extensions to tomographic schemes that consider unknown parameters in both state and process parameters may also be possible \cite{Branczyk2012a,Mogilevtsev2012,Merkel2012,Blume-Kohout2013,Takahashi2013}. However, we anticipate the nonlinear relationship between measurement statistics and unknown parameters to provide additional challenges. 

Our figure of merit is closely related to the trace distance, introduced by Scott \cite{Scott2006}. However, the trace distance is only applicable for signal-independent noise, while our figure of merit accounts for signal-dependent noise. Our figure of merit does not posses the same operational interpretation that might make other figures of merit---such as the fidelity \cite{deBurgh2008}---more appealing, however it is vastly simpler to work with as it does not require computational averaging over input states. 

We anticipate that the method introduced in this paper will help the design of tomographic experiments performed in laboratories with minimal available resources. Furthermore, our figure of merit can be used to compare the performance of different tomographic protocols, in complement to other figures of merit such as the fidelity.

\section{Acknowledgements}
We thank Daniel James for helpful discussions. We also thank Joe Altepeter, Hubert de Guise and Alessandro Fedrizzi for helpful comments on the manuscript. MM acknowledges support from NSERC (USRA) and The Fields Institute. AMB acknowledges support from DARPA (QuBE). 

\appendix

\section{Six-measurement example for qubits}\label{sec:example}

Consider a qubit encoded in the polarization of a single photon. One of the simplest measurements one can implement is a two-outcome POVM with elements 
\begin{align}\label{eq:ph1}
\hat{E}_1={}&\mathcal{N}\ket{H}{}\bra{H}{};~~~\hat{E}_2={}\mathcal{N}\ket{V}{}\bra{V}{}\,.
\end{align}
In fact, these operators correspond to the special case of a projective-value measure (PVM), where $\hat{E}_i\hat{E}_{j}=\hat{E}_i\delta_{i,j}$. This  measurement is implemented by sending multiple copies of the state through a polarizing beam splitter (which splits the state into horizontal and vertical modes) and counting the number of photons in each output port. 

However, this is not enough to reconstruct the quantum state because this PVM isn't informationally complete. Two additional PVMs can be constructed by inserting a half-wave plate (HWP) followed by a quarter-wave plate (QWP) in front of the PBS. Setting the HWP and QWP at $22.5^{\circ}$ and $45^{\circ}$ to the optical axis, respectively, generates the following PVM:
\begin{align}\label{eq:ph2}
\hat{E}_3={}&\mathcal{N}\ket{D}{}\bra{D}{};~~~\hat{E}_4={}\mathcal{N}\ket{A}{}\bra{A}{}\,,
\end{align}
where $\ket{D/A}{}=\ket{H}{}\pm \ket{V}{}/\sqrt{2}$. Setting them to $0^{\circ}$ and $45^{\circ}$ gives
\begin{align}\label{eq:ph2}
\hat{E}_5={}&\mathcal{N}\ket{R}{}\bra{R}{};~~~\hat{E}_6={}\mathcal{N}\ket{L}{}\bra{L}{}\,,
\end{align}
where $\ket{R/L}{}=(\ket{H}{}\pm i\ket{V}{})/\sqrt{2}$. For all three PVMs, we have assumed equal data collections times and therefore the total number of counts is $\mathcal N$ in all three cases. 

 Performing the experiment, we get a set of six numbers which correspond to the counts at both output ports of the PBS for each of the three PVMs. These can be written as 
\begin{align}
\mathbf{n}={}&(n_{1}, n_{2},n_{3}, n_{4},n_{5}, n_{6})^{\mathrm{T}}\,.
\end{align}

We are interested in finding the vector
\begin{align}
\mathbf{a}={}&({S}_{0}, {S}_{1}, {S}_{2}, {S}_{3} )^T\,,
\end{align}
which characterizes the state of our qubit.

The measurement matrix $\mathbf{M}$, which relates $\mathbf{n}$ to $\mathbf{a}$ via Eq. (\ref{eq:matP}), is given by
\begin{align}
\mathbf{M}=\frac{\mathcal{N}}{2}\left(\begin{array}{cccc}1 & 0 & 0 & 1 \\1 & 0 & 0 & -1 \\1 & 1 & 0 & 0 \\1 & -1 & 0 & 0 \\1 & 0 & 1 & 0 \\1 & 0 & -1 & 0\end{array}\right)\,,
\end{align}
where
\begin{align}
\mathbf{M}_{i, j}= \frac{1}{2}\mathrm{Tr}[ \hat \sigma_{j-1} \hat{E}_{i}] \,,
\end{align}
and $\hat\sigma_0=\ket{H}{}\bra{H}{}+\ket{V}{}\bra{V}{}$, $\hat\sigma_1=\ket{H}{}\bra{V}{}+\ket{V}{}\bra{H}{}$, $\hat\sigma_2=i\ket{V}{}\bra{H}{}-i\ket{H}{}\bra{V}{}$ and $\hat\sigma_3=\ket{H}{}\bra{H}{}-\ket{V}{}\bra{V}{}$. The pseudoinverse of $\mathbf{M}$ is given by
\begin{align}
\mathbf{M}^{+}=\frac{1}{\mathcal{N}}\left(
\begin{array}{cccccc}
 \frac{1}{3} & \frac{1}{3} & \frac{1}{3} & \frac{1}{3} & \frac{1}{3} &
   \frac{1}{3} \\
 0 & 0 & 1 & -1 & 0 & 0 \\
 0 & 0 & 0 & 0 & 1 & -1 \\
 1 & -1 & 0 & 0 & 0 & 0 \\
\end{array}
\right)\,,
\end{align}
which can be used to determine the vector $\mathbf{a}$ as follows:
\begin{align}
\mathbf{a}=\mathbf{M}^+\mathbf{n}=\left(
\begin{array}{c}
 \frac{1}{3}\sum_{i=1}^6 n_{i} \\
 n_{3}-n_{4} \\
 n_{5}-n_{6} \\
 n_{1}-n_{2} \\
\end{array}
\right)\,.
\end{align}
The vector $\mathbf{a}$ then gives the density matrix according to
\begin{eqnarray}
\hat \rho=\frac{1}{2 }  \sum_{{j}=1}^{4} \mathbf{a}_{j} \hat \sigma_{j-1} \,.
\end{eqnarray}

\section{Discussion of linear inversion and positive semi-definiteness}\label{sec:pos}

Use of linear inversion (LI) for state reconstruction yields coefficients $S_i$ that may not necessarily correspond to a physical state, i.e., the estimated density matrices may be unnormalized or negative definite. 

We can denote the estimated coefficients obtained from this method by $\vec{S}_{\mathrm{LI}} = (S_{0, \mathrm{LI}}, S_{1, \mathrm{LI}}, ..., S_{(d^2 -1), \mathrm{LI}} )$. We also denote coefficients corresponding to the actual state (A) of the system by $\vec{S}_{\mathrm{A}} = (S_{0, \mathrm{A}}, S_{1, \mathrm{A}}, ..., S_{(d^2 -1), \mathrm{A}} )$. 

The proposed figure of merit, the EWV, is the expectation value of $\sum_{i=0}^{d^2-1} (S_{i, \mathrm{LI}} - S_{i, \mathrm{A}})^{2}$. As $\vec{S}_{\mathrm{LI}} $ may not correspond to a physical state, the experimenter could solve this problem by, say, looking for a physical state  (Ph) that is closest to the one obtained by linear inversion; we denote this state by $\vec{S}_{\mathrm{Ph}}$ and characterize the closeness of states by the distance function $d (\vec{S_{2}}, \vec{S_{1}}) = (\sum_{i=0}^{d^2-1} (S_{i, 2} - S_{i, 1})^{2})^{1/2}$. In this scenario, we are interested in the expectation value of $d^{2} (\vec{S}_{\mathrm{Ph}}, \vec{S}_{\mathrm{A}})$,  as a measure for the error. Since $\vec{S}_{\mathrm{Ph}}$ is the closest physical state to $\vec{S}_{\mathrm{LI}}$, we have $ d (\vec{S}_{\mathrm{Ph}}, \vec{S}_{\mathrm{LI}}) \leq d (\vec{S}_{\mathrm{A}}, \vec{S}_{\mathrm{LI}})$. This, along with the triangle inequality, yields $d (\vec{S}_{\mathrm{Ph}}, \vec{S}_{\mathrm{A}}) \leq d (\vec{S}_{\mathrm{Ph}}, \vec{S}_{\mathrm{LI}}) + d (\vec{S}_{\mathrm{LI}}, \vec{S}_{\mathrm{A}}) \leq 2 d (\vec{S}_{\mathrm{LI}}, \vec{S}_{\mathrm{A}}) $. 

The error computed while taking the positive semi-definiteness of the density matrix into account would be bounded by a coefficient of the old figure of merit. Therefore, minimizing the proposed figure of merit, EWV, also minimizes an upper bound for the error that takes positive semi-definiteness into account.

We add that the geometry of physical states is especially complicated and there are many open problems---e.g. the existence of SIC-POVMs---associated with it.  We therefore consider our approach justified in that it provides a simple procedure that leads to reasonable results. 

\section{Simplify expression for equally-weighted variance}\label{sec:simpI}

In this section, we detail the steps between Eqs. (\ref{eq:something}) and (\ref{eq:im}). For reference, we repeat Eq.  (\ref{eq:something}) here:

\begin{align}\label{eq:test}
\begin{split}
\mathrm{EWV}={}& \sum_{{i},{k}=1}^{d^{2}}\sum_{j=1}^{N} \left(\mathbf{M}^{+}_{{i}{j}}\right)^2\mathbf{M}_{jk}\mathbf{a}_{k}\,.
\end{split}
\end{align}

We define a vector
\begin{align}
\mathbf{g}_{k}={}&\sum_{i=1}^{d^{2}}\sum_{{j}=1}^{N} \left(\mathbf{M}^{+}_{{i}{j}}\right)^2\mathbf{M}_{{j}{k}}\,.
\end{align}
such that Eq. (\ref{eq:test}) becomes
\begin{align}\label{eq:dfsd}
\begin{split}
\mathrm{EWV}={}& \sum_{{k}=1}^{d^{2}}\mathbf{g}_{k}\mathbf{a}_{k}\,.
\end{split}
\end{align}

We rewrite the term $\mathbf{g}_{k}$ as
\begin{align}
\mathbf{g}_{k}\equiv {}&\mathrm{Tr}\left[\mathbf{M}^{+}\mathbf{Y}^{(k)}(\mathbf{M}^{+})^{\mathrm{T}}\right]
\end{align}
where $\mathbf{Y}^{(k)}$ is an $N\times N$ diagonal matrix whose diagonal elements are the elements of the $k$th column of $\mathbf{M}$, i.e. $\mathbf{Y}_{jj}^{(k)}=\mathbf{M}_{{j}{k}}$ . We perform a singular value decomposition $\mathbf{M}^{+}=(\mathbf{V}^+)^{\mathrm{T}}\mathbf{S}^+\mathbf{U}^+$ (i.e. $\mathbf{M}=\mathbf{U}\mathbf{S}\mathbf{V}^{\mathrm{T}}$),  where $\mathbf{U}$ and $\mathbf{V}$ are   $d^2\times d^2$ and $N\times N$ matrices recpectively, to give
\begin{align}
\mathbf{g}_{k}={}&\mathrm{Tr}\left[(\mathbf{U}^+)^{\mathrm{T}}(\mathbf{S}^+)^{\mathrm{T}}\mathbf{V}^+(\mathbf{V}^+)^{\mathrm{T}}\mathbf{S}^+\mathbf{U}^+\mathbf{Y}^{(k)}\right]\\
={}&\mathrm{Tr}\left[(\mathbf{S}^+)^{\mathrm{T}}\mathbf{S}^+\mathbf{X}\right]
\end{align}
where $\mathbf{X}=\mathbf{U}^+\mathbf{Y}^{(k)}(\mathbf{U}^+)^{\mathrm{T}}$. Since $\mathbf{S}^+$ is a diagonal matrix of inverse singular values of $\mathbf{M}$, i.e. $\mathbf{S}^+_{i}=1/\mu_{i}$, we can write
\begin{align}
\mathbf{g}_{k}={}&\sum_{{i}=1}^{d^2}\frac{1}{\mu_{i}^2}\mathbf{X}_{{i},{i}}\\
={}&\sum_{{i}=1}^{d^2}\frac{1}{\mu_{i}^2}\sum_{j=1}^{d^2}(\mathbf{U}^+_{{i}{j}})^2\mathbf{Y}_{jj}^{(k)}\\\label{eq:blah}
={}&\sum_{{i}=1}^{d^2}\frac{1}{\mu_{i}^2}\sum_{j=1}^{d^2}(\mathbf{U}^+_{{i}{j}})^2\mathbf{M}_{{j}{k}}
\end{align}

Inserting Eq. (\ref{eq:blah}) into Eq. (\ref{eq:dfsd}), we now have
\begin{align}\label{eq:dgdfs}
\begin{split}
\mathrm{EWV}={}& \sum_{{i}=1}^{d^2}\frac{1}{\mu_{i}^2}\sum_{{k}=1}^{d^{2}}\mathbf{r}_{{i},{k}}\mathbf{a}_{k}\,.
\end{split}
\end{align}
where
\begin{align}\label{eq:dgdfsdf}
\mathbf{r}_{{i},{k}}={}&\sum_{j=1}^{d^2}(\mathbf{U}^+_{{i}{j}})^2\mathbf{M}_{{j}{k}}\,.
\end{align}

Eqs.  (\ref{eq:dgdfs}) and (\ref{eq:dgdfsdf}) correspond to Eqs.  (\ref{eq:im}) and (\ref{eq:r}) in the main text.

\section{SIC-POVM}\label{sec:sicpovm}

In this section, we show that the average equally-weighted variance $\left < \mathrm{EWV} \right >$ for any symmetric informationally-complete POVM (SIC-POVM) necessarily reaches the lower bound $\left < \mathrm{EWV} \right >_{\mathrm{LB}}$. 

SIC-POVMs are therefore optimal for QST in the sense that they are the least sensitive POVM to Poisson noise. This is consistent with the result that SIC-POVMs are the most immune minimal informationally complete POVM for signal-independent noise \cite{Scott2006}. We note, however, that the existence of SIC-POVMs in arbitrary dimensions has not yet been proven or disproven. 

To show that $\left < \mathrm{EWV} \right >$ for SIC-POVMs equals $\left < \mathrm{EWV} \right >_{\mathrm{LB}}$, we show that the eigenvalues of the matrix $\mathbf{M}^{\mathrm{T}}\mathbf{M}$ (constructed using a SIC-POVM) are $\mu_{1}^{2}=\alpha^{2}$ and $\mu_{2}^{2}=\dots=\mu_{d^{2}}^{2}=\alpha^{2}/(d+1) $, i.e. the eigenvalues for the lower bound.  

First, consider an unnormalized SIC-POVM given by $\{ \hat{E}_{j}  \}_{j=1}^{d^{2}}$. By definition
\begin{subequations}\label{eq:sicpovm}
\begin{align} 
\mathrm{Tr}[\hat{E}_{i} \hat{E}_{j}] ={}&\mathcal{N}^{2} \, \mathrm{Tr}[\hat {F}_{i} \hat {F}_{j}] \,;\\
={}&\left\{\begin{array}{ccc}\frac{\mathcal{N}^2}{d^{2}(d+1)} & ~\mathrm{for}~ & i \neq j \\
\frac{\mathcal{N}^2}{d^{2}} & ~\mathrm{for}~ & i = j\end{array}\right.
\end{align}
\end{subequations}
where $F_{i}=\hat{E}_{i} / \mathcal{N}$ are normalized probability operators. One can also see that:
\begin{align}
\alpha=\mathrm{Tr}[\hat{E}_{i}]=\mathcal{N}\mathrm{Tr}[\hat F_{i}] =\frac{\mathcal{N}}{d}\,.
\end{align} 

Using the fact that the set of operators $\mathbf{B}$, defined in Eq. (\ref{eq:basis}), form an orthonormal basis in the sense of the Hilbert-Schmidt inner product, we write
\begin{align} \label{eq:innerrows}
\begin{split}
\mathrm{Tr}[\hat{E}_{i} \hat{E}_{j}] =
 \frac{1}{2} \sum_{{k}={0} } ^{d^{2}-1} \mathrm{Tr}[ \hat \sigma_{k} \hat{E}_{i}] \mathrm{Tr}[ \hat \sigma_{k} \hat{E}_{j}] .
\end{split}
\end{align}
We can equate Eqs. (\ref{eq:sicpovm})  and (\ref{eq:innerrows}) to give
\begin{align}\label{eq:eq}
 \sum_{{k}={0} } ^{d^{2}-1} \mathrm{Tr}[ \hat \sigma_{k} \hat{E}_{i}] \mathrm{Tr}[ \hat \sigma_{k} \hat{E}_{j}] ={}&\left\{\begin{array}{ccc}\frac{2\mathcal{N}^2}{d^{2}(d+1)} & ~\mathrm{for}~ & i \neq j \\
\frac{2\mathcal{N}^2}{d^{2}} & ~\mathrm{for}~ & i = j\end{array}\right.\,.
\end{align}

Now consider the inner product between two rows of the matrix $\mathbf{M}$. For reference, we rewrite Eq. (\ref{eq:column}) here: 
\begin{align}\label{eq:column2}
\mathbf{M}_{i, j}= \frac{1}{\sqrt{2d}}\mathrm{Tr}[ \hat \sigma_{j-1} \hat{E}_{i}] \,.
\end{align}
From Eqs. (\ref{eq:eq}) and (\ref{eq:column2}), we can see that 
\begin{align}\label{eq:rows}
\mathrm{row}_{i} \left ( \mathbf{M} \right )  \cdot \mathrm{row}_{j} \left ( \mathbf{M}\right ) = {}&\left\{\begin{array}{ccc}\frac{\mathcal{N}^2}{d^{3}(d+1)} & ~\mathrm{for}~ & i \neq j \\
\frac{\mathcal{N}^2}{d^{3}} & ~\mathrm{for}~ & i = j\end{array}\right.\,.
\end{align}
where $\cdot$ is the inner product in the usual sense. 

The next piece of the puzzle is to prove that $\mathbf{M}^{\mathrm{T}}\mathbf{M}$ is diagonal; the diagonal entries will therefore be the eigenvalues of  $\mathbf{M}^{\mathrm{T}}\mathbf{M}$. To do so, we use Eqs. (\ref{eq:column2}) and (\ref{eq:rows}) to construct the following matrix:
\begin{align}\label{eq:magicmatrix}
\mathbf{O}=\frac{d\sqrt{d+1}}{\mathcal{N}} \mathbf{L}\,,
\end{align}
where the matrix $\mathbf{L}$ is equal to $\mathbf{M}$ with its first column $( \mathcal{N} / d^2, \dots,\mathcal{N} / d^2 )$ divided by $\sqrt{(d+1)}$. It is not difficult to confirm that the rows of  $\mathbf{O}$ are orthonormal. Thus, $\mathbf{O}$ is an orthogonal matrix and consequently the columns of $\mathbf{O}$ are also orthonormal. We can therefore conclude that the columns of $\mathbf{M}$ are orthogonal, i.e.
\begin{align}\label{eq:columnsproduct}
\mathrm{col}_{i} \left ( \mathbf{M} \right )  \cdot \mathrm{col}_{j} \left ( \mathbf{M}\right ) = 0    ~~\mathrm{for}~~  i \neq j \,,
\end{align}
and that $\mathbf{M}^{\mathrm{T}}\mathbf{M}$ is a diagonal matrix. 

We now want to calculate expressions for ${d}_{i}$. From Eq. \ref{eq:column2}, we see that
\begin{align}\label{eq:column3}
d_{1}=\mathrm{col}_{1} \left ( \mathbf{M} \right )  \cdot \mathrm{col}_{1} \left ( \mathbf{M}\right ) = \frac{\mathcal{N}^{2}}{d^{2}} \,.
\end{align} 
From the  orthogonality of $\mathbf{O}$ we also see that
\begin{align}\label{eq:columnsproduct2}
d_{i}=\mathrm{col}_{i} \left ( \mathbf{M} \right )  \cdot \mathrm{col}_{i} \left ( \mathbf{M}\right ) = \frac{\mathcal{N}^{2}}{d^{2}(d+1)}   ~~\mathrm{for}~~  i > 1\,. \end{align}

By showing that  $\mathbf{M}^{\mathrm{T}}\mathbf{M}$ is a diagonal matrix with the diagonal entries $d_{1}=\mathcal{N}^{2} /d^{3} $ and $d_{2}=\dots=d_{d^{2}}=\mathcal{N}^{2} /(d^{3}(d+1))  $, we show that the eigenvalues of $\mathbf{M}^{\mathrm{T}}\mathbf{M}$ are $\mu_{1}^{2}=\alpha^2$ and $\mu_{2}^{2}=\dots=\mu_{d^{2}}^{2}=\alpha^{2} /(d+1) $, i.e. the eigenvalues for the lower bound for $N=d^2$ in Eqs. (\ref{eq:singlenoise}).

We therefore prove that SIC-POVMs do reach the lower bound. We now need to prove that \emph{only} SIC-POVMs reach the lower bound, when $N=d^{2}$.

We do so by showing that any set of operators $\{ {\hat\Pi}_{i} \} _{i=1}^{N}$, where $N=d^2$, that reaches the lower bound will necessarily correspond to a SIC-POVM. From the definition of the SVD in Eq. (\ref{eq:113}) and the expressions for the singular values in Eqs. (\ref{eq:minimumsingular}) and (\ref{eq:minimumsingular2}), we can write
\begin{align} \label{eq:eigenvector}
\mathbf{M}^{\mathrm{T}}\mathbf{M}= \frac{\mathcal{N}^{2}}{d^{2}} \bf{V} \left(\begin{array}{ccccc}1 & 0 & . & . & 0 \\0 & \frac{1}{d+1} & . & . & 0 \\0 & 0 & . & . & 0 \\0 & 0 & . & . & 0 \\0 & 0 & . & . & \frac{1}{d+1}\end{array}\right)\bf{V}^{\mathrm{T}}\,,
\end{align}
where $\bf V$ is a $d^{2} \times d^{2}$ orthogonal matrix. Eq. (\ref{eq:eigenvector}) takes this form for any POVM that minimized the noise. 

Since the highest eigenvalue of $\mathbf{M}^{\mathrm{T}}\mathbf{M}$ is $\mathcal{N}^{2}/d^{2}$, we can write the following for any $d^{2}$-dimensional vector of unit length, $\mathbf{v}$:
\begin{align}
{\mathbf{v}} ^{\mathrm{T}} \mathbf{M}^{\mathrm{T}}\mathbf{M}\mathbf{v} \leq \frac{\mathcal{N}^{2}}{d^{2}}\,.
\end{align}
One can check that the equality holds for the vector $\mathbf{v}_{1}={(1, 0, 0, \dots, 0)}^{\mathrm{T}}$ and, therefore, $\mathbf{v}_{1}$ is the eigenvector of $\mathbf{M}^{\mathrm{T}}\mathbf{M}$ with eigenvalue $\mathcal{N}^{2}/d^{2}$. It follows that $\mathbf{M}^{\mathrm{T}}\mathbf{M}$ has the form
\begin{align}\label{eq:V}
\mathbf{M}^{\mathrm{T}}\mathbf{M} =\frac{\mathcal{N}^{2}}{d^{2}} \left ( {\mathbf{v}_{1}} {\mathbf{v}_{1}}^{\mathrm{T}} + \frac{1}{d+1} \left ( \mathbb {I}_{d} - {\mathbf{v}_{1}} {\mathbf{v}_{1}}^{\mathrm{T}}  \right ) \right)\,,
\end{align}
and that
\begin{align} \label{eq:diagonal}
\mathbf{M}^{\mathrm{T}}\mathbf{M}= \frac{\mathcal{N}^{2}}{d^{2}} \left(\begin{array}{ccccc}1 & 0 & . & . & 0 \\0 & \frac{1}{d+1} & . & . & 0 \\0 & 0 & . & . & 0 \\0 & 0 & . & . & 0 \\0 & 0 & . & . & \frac{1}{d+1}\end{array}\right) .
\end{align}

From the diagonal entries of $\mathbf{M}^{\mathrm{T}}\mathbf{M}$, we see that Eq. (\ref{eq:columnsproduct2}) holds. Since $\mathbf{M}^{\mathrm{T}}\mathbf{M}$ is diagonal, Eq .(\ref{eq:columnsproduct}) is also true.

Working backwards, we can see that matrix $\mathbf{O}$, defined in Eq. (\ref{eq:magicmatrix}), is an orthogonal matrix, as it consists of orthonormal columns. From this, one can verify Eq. (\ref{eq:rows}) and consequently (\ref{eq:sicpovm}).  The measurement matrix $\mathbf{M}$ that reaches the lower bound for the set of measurements $\{{\hat\Pi}_{i} \} _{i=1}^{d^{2}}$, will necessarily correspond to a SIC-POVM.

To  calculate the uncertainty due to a Poisson source, we worked with the number of counts, rather than the corresponding probabilities (as was done in \cite{Scott2006}). Experimentally, probabilities are obtained  by dividing the number of counts by the total number of counts, which is the summation of number counts in a POVM. Thus, probabilities are dependent parameters. This would make calculation of uncertainties a difficult task when dealing with the probabilities. In the presence of signal-dependent noise, all probabilities obtained from a POVM have the same uncertainty and results obtained when working with counts reduce to those obtained when working with probabilities.

\end{document}